\documentclass[a4paper,fleqn]{cas-dc}

\usepackage[authoryear]{natbib}
\usepackage[version=3]{mhchem}
\usepackage{lineno}
\usepackage{setspace}

\def\tsc#1{\csdef{#1}{\textsc{\lowercase{#1}}\xspace}}
\tsc{WGM}
\tsc{QE}
\tsc{EP}
\tsc{PMS}
\tsc{BEC}
\tsc{DE}

\begin{document}
\let\WriteBookmarks\relax
\def\floatpagepagefraction{1}
\def\textpagefraction{.001}
\shorttitle{Exo-planetary post-nebula volatilisation}
\shortauthors{Harrison et~al.}

\title [mode = title]{Evidence for post-nebula volatilisation in an exo-planetary body}

\author[1]{John H. D. Harrison}[orcid=0000-0002-9971-4956]
\cormark[1]
\ead{jhdh2@cam.ac.uk}
\author[1,2]{Oliver Shorttle}
\author[1]{Amy Bonsor}

\address[1]{Institute of Astronomy, University of Cambridge, Madingley Road, Cambridge, CB3 0HA, UK}

\address[2]{Department of Earth Sciences, University of Cambridge, Downing Street, Cambridge, CB2 3EQ, UK}



\begin{abstract}
The loss and gain of volatile elements during planet formation is key for setting their subsequent climate, geodynamics, and habitability. Two broad regimes of volatile element transport in and out of planetary building blocks have been identified: that occurring when the nebula is still present, and that occurring after it has dissipated. Evidence for volatile element loss in planetary bodies after the dissipation of the solar nebula is found in the high Mn to Na abundance ratio of Mars, the Moon, and many of the solar system's minor bodies. This volatile loss is expected to occur when the bodies are heated by planetary collisions and short-lived radionuclides, and enter a global magma ocean stage early in their history. The bulk composition of exo-planetary bodies can be determined by observing white dwarfs which have accreted planetary material. The abundances of Na, Mn, and Mg have been measured for the accreting material in four polluted white dwarf systems. Whilst the Mn/Na abundances of three white dwarf systems are consistent with the fractionations expected during nebula condensation, the high Mn/Na abundance ratio of GD362 means that it is not (>3$\sigma$). We find that heating of the planetary system orbiting GD362 during the star's giant branch evolution is insufficient to produce such a high Mn/Na. We, therefore, propose that volatile loss occurred in a manner analogous to that of the solar system bodies, either due to impacts shortly after their formation or from heating by short-lived radionuclides. We present potential evidence for a magma ocean stage on the exo-planetary body which currently pollutes the atmosphere of GD362.
\end{abstract}


\maketitle

\section{Introduction}

Most rocky planet forming material originates in protoplanetary discs, when hot circumstellar gas condenses into solid matter at the midplane of the disc \citep{Williams2011}. Through the subsequent growth of dust particles into ever larger aggregates: pebbles, planetary embryos and eventually terrestrial planets are formed. The bulk composition of both Earth and the chondrites are well explained using such a nebula condensation model, despite Earth's complex and ongoing differentiation history and alteration processes having affected chondrite parent bodies \citep{Anders1964, Wasson1988, Lodders2003, Lodders2010, Siebert2018}. The refractory elemental abundances of the chondrites match those of the Sun, whilst their volatile elemental abundances are depleted relative to the Sun in accordance with their individual elemental condensation temperatures \citep{McDonough1995, McDonough2003, Lodders2003}. Different thermal conditions in the solar nebula have imparted a fundamental compositional fingerprint on planetary material through condensation processes. How universal this process is to planet formation is a key question, one which can be answered through study of extra-solar rocky material.
\newline

It is possible to probe the abundances of planetary material from outside the solar system by observing metal features in the atmospheres of white dwarfs \citep{JuraYoung2014}. White dwarfs are the faint remnants of the cores of stars like the Sun, and theoretically they should have atmospheres only composed of hydrogen and helium \citep{Althaus2010, Koester2013}. The metal features observed in many white dwarfs are thought to be present due to the accretion of rocky planetary bodies \citep{JuraYoung2014, Farihi_review}. As the properties of the metal spectral features can be used to constrain the relative abundances of the metals in white dwarf atmospheres' \citep{Koester09}, polluted white dwarfs can probe the composition of exo-planetary bodies and test whether nebula condensation can explain the abundances found \citep{Harrison2018}.
\newline

Nebula condensation is not the sole process which determines the composition of rocky planetary material. The composition of bodies can also be significantly altered post formation \citep{ONeillPalme2008}. Significant melting and the formation of a global magma ocean can occur on rocky planetary bodies due to high energy planetary impacts and the decay of short-lived radioactive isotopes \citep{Keil2000, Pringle2014, Wang2016, Hin2017, Siebert2018}. This heating, referred to in this work as post-nebula volatilisation, causes the preferential loss of volatile elements, especially on less massive bodies which do not have sufficient surface gravity to stop the thermal escape of the vapour \citep{ONeillPalme2008, Pringle2014}. Post-nebula volatilisation occurs at much higher pressures and in much more oxidising conditions than solar nebula condensation \citep{Visscher2013}. Thus, individual elemental behaviours and speciations may be significantly different between these two regimes, and consequently the abundance signatures created during post-nebula volatilisation need not match those expected from nebula condensation \citep{Sossi2018}. The atmospheric compositions of polluted white dwarfs could, therefore, potentially provide evidence for post-nebula volatilisation in exo-planetary bodies. In the case of white dwarf systems there is the added complexity that the planetary bodies must survive the star's giant branch evolution, so may experience strong heating processes as the host star leaves the main sequence and increases in luminosity.
\newline

As well as trends related to volatility, planetary compositions can be altered by large scale melting and the segregation of siderophilic elements into a planetary core. Collisions between differentiated bodies can separate core and mantle material and can lead to the production of bodies with bulk compositions dissimilar to those of the bodies that condensed out of the stellar nebula. Iron meteorites and the achondrites are examples which provide evidence for the occurrence of this process in the solar system \citep{Scott1975,Lodders1998,FeMeteorites,Asteroids4}. As planetary differentiation preferentially moves siderophilic elements into the core and lithophilic elements into the mantle and crust, planetary bodies are no longer homogeneous, thus, disruptive collisions may produce fragments which are correspondingly enhanced/depleted in such elements. The atmospheric compositions of polluted white dwarfs have been used to provide evidence of planetary differentiation and collisional processing in exo-planetary systems due to observations of siderophile/lithophile rich/poor atmospheric compositions \citep{JuraYoung2014, Harrison2018}.
\newline

\begin{figure}
	\centering
		\includegraphics[width=\columnwidth]{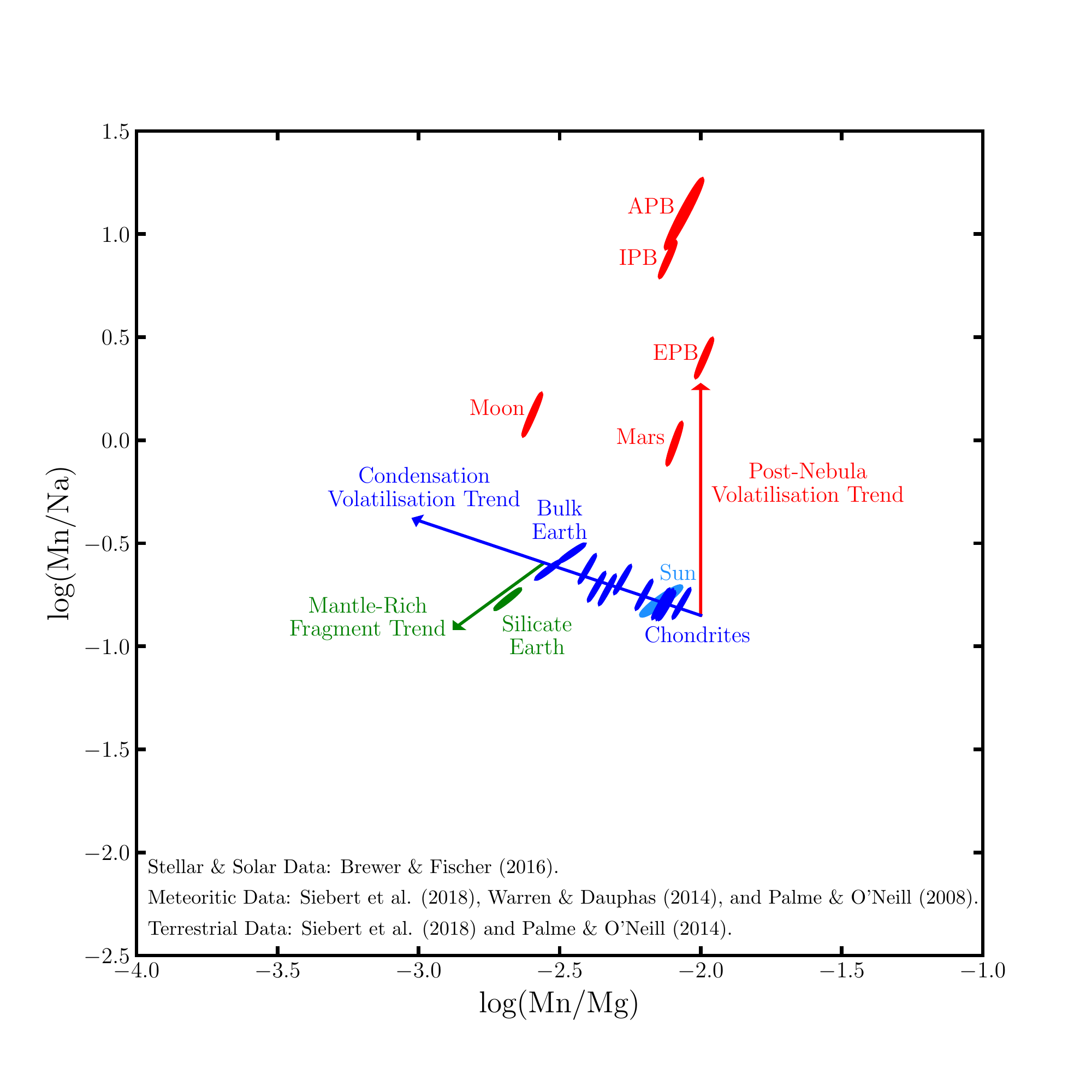}
	\caption{A modified version of Figure 1 from \protect\cite{Siebert2018}. The Mn/Na and Mn/Mg ratios of solar system bodies and fragments can be explained by three processes: condensation from the solar nebula (blue), planetary differentiation (green), and post-nebula volatilisation (red). The three coloured arrows indicate trends relating to the three processes based on starting with initially solar abundance values. Errors are displayed as 1$\sigma$ error ellipses to capture the correlation between the axes.}
	\label{FIG:1}
\end{figure}

We can trace the environment of volatilisation processes using Mn, Na, and Mg. Figure \ref{FIG:1} is a modified version of Figure 1 from \cite{Siebert2018} (Data sources are \cite{ONeillPalme2008,Dauphas2014,Palme2014,FischerBrewer2016,Siebert2018}). The positions of the solar system bodies in log(Mn/Na) and log(Mn/Mg) space can be readily explained by three processes: condensation from the solar nebula (blue), planetary differentiation (green), and post-nebula volatilisation (red).
\newline

Mn can behave as both a lithophile and a siderophile while Na and Mg dominantly behave as lithophiles, hence, the Mn/Na ratio and the Mn/Mg ratio of a body can be altered by differentiation, collisions, and fragmentation. Evidence for this is found in their abundances in the silicate Earth where Mn is depleted relative to bulk Earth (i.e., partially lost to the core), while Na and Mg are enriched (i.e., retained in the mantle)
 \citep{McDonough2003,Palme2014,Siebert2018}. The green vector on Figure \ref{FIG:1} corresponds to the abundances expected for increasingly mantle-rich collisional fragments of differentiated bodies.
\newline

Mn and Na are both volatile elements in solar nebula conditions with 50\% condensation temperatures of 1158\,K and 958\,K respectively, while Mg is a non-volatile element with a 50\% condensation temperature of 1336\,K \citep{Lodders2003}. Therefore, bodies which experienced hotter formation temperatures are expected to have higher Mn/Na ratios and lower Mn/Mg ratios. Evidence for this is found in the abundances of the chondrites and bulk Earth \citep{ONeillPalme2008, Siebert2018} which lie along the blue condensation vector plotted on Figure \ref{FIG:1}.
\newline

However, in detail the relative volatility of Mn and Na is heavily dependent on the oxygen fugacity and the pressure at which the condensation/volatilisation process is occurring. Impact generated silicate melting (post-nebula volatilisation) would have occurred under much more oxidising conditions and at much higher pressures than nebula condensation \citep{Visscher2013,ONeillPalme2008,Siebert2018}. Such conditions would cause Na to become much more volatile relative to Mn and Mg, and thus, would cause preferential loss of Na, leading to enhanced Mn/Na ratios but unchanged Mn/Mg ratios \citep{ONeillPalme2008, Pringle2014}. Evidence for post-nebula volatilisation is found in the super-chondritic Mn/Na ratios of Mars, the Moon, the Angrite parent body, the Ibitira parent body, and the Eucrite parent body. This process is described by the composition of objects moving along the red vector in Figure \ref{FIG:1}. 
\newline

By finding the Mn, Mg, and Na abundances of rocky bodies in exo-planetary systems we can attempt to answer two main questions: Firstly, do the three processes of nebula condensation, differentiation, and post-nebula volatilisation occur regularly in other planetary systems? Secondly, are these three processes the major factors which determine the bulk composition of rocky exo-planetary bodies?
\newline

In this work we use the Mn/Na ratio and the Mn/Mg ratio of the exo-planetary bodies which pollute white dwarfs to assess whether the three processes known to alter the Mn/Na ratio and Mn/Mg ratio in the solar system can explain the abundances observed. We also investigate whether the effect of post-main sequence stellar evolution is expected to alter the composition of the rocky bodies polluting white dwarfs. In section 2 we outline the polluted white dwarf data used, in section 3 we outline the post-main sequence heating model used, in section 4 we discuss the caveats of our work and our results, and in section 5 we state our conclusions.

\section{Polluted white dwarf data} \label{PWDD}

Currently the most direct method for measuring the bulk composition of rocky bodies in exo-planetary systems is by observing the atmospheres of externally polluted white dwarfs. Externally polluted white dwarfs are cool white dwarf stars with metal features in their spectra \citep{JuraYoung2014}. Metal absorption lines have been detected in more than one thousand cool white dwarfs \citep{Kepler2016, Hollands2017, Coutu2019}. The polluting metals must have been accumulated in the upper atmospheres of the cool white dwarf stars relatively recently because the cooling ages of the white dwarfs (of the order tens of millions of years to billions of years) are far longer than the time it takes for the metals to sink out of the upper atmosphere and become unobservable (of the order days to millions of years) \citep{Koester09,JuraYoung2014}. For the white dwarf stars in question the polluting metals cannot originate from the interstellar medium, the fallback of the star's giant branch winds, or the radiative levitation of primordial metals \citep{Farihi10ism,JuraYoung2014,Farihi_review, Veras_review,Preval2019}. It is now widely accepted that for these stars the polluting material is of an exo-planetary origin, and therefore, by measuring the relative abundances of the metals in the polluted white dwarf atmospheres a unique insight into the bulk compositions of exo-planetary rocky material can be found \citep{JuraYoung2014,Farihi_review, Veras_review}.
\newline

\begin{table}[width=\linewidth,cols=4,pos=h]
\caption{The atmospheric abundances (log number fraction) for the four polluted white dwarf stars with Mg, Na, and Mn abundance measurements. The atmospheric abundances were derived in \protect\cite{Zuckerman07,Dufour2012,Xu2013,Hollands2017,Swann2019}.}
\label{tbl1}
\begin{tabular*}{\tblwidth}{@{} LLLLLL@{} }
\toprule
System & log(Mg/He) & log(Na/He) & log(Mn/He)\\
\midrule
GD362 & $-5.98 \pm 0.25$ & $-7.79 \pm 0.20$ & $-7.47 \pm 0.10$ \\
J0738+1835 & $-4.68 \pm 0.07$ & $-6.36 \pm 0.16$ & $-7.11 \pm 0.10$ \\
WD\,0446-255 & $-6.60 \pm 0.10$ & $-7.90 \pm 0.10$ & $-9.10 \pm 0.10$ \\
J1535+1247 & $-7.36 \pm 0.10$ & $-8.72 \pm 0.05$ & $-9.80 \pm 0.20$ \\
\bottomrule
\end{tabular*}
\end{table}

Table 1 contains the atmospheric abundances for the four polluted white dwarfs investigated in this work. The atmospheric abundances were derived in \cite{,Zuckerman07,Dufour2012,Xu2013,Hollands2017,Swann2019} (the Mn value was not published in \cite{Hollands2017} as that paper focused on a restricted set of elements, however, the Mn abundance was provided by Mark Hollands via private communication) and they are currently the only white dwarfs which have measured abundances of Mn, Mg, and Na in their atmospheres. Six other white dwarfs are known to have two of the three elements of interest in their atmospheres, however, with only upper limits at best on the third elemental abundance we do not investigate these white dwarfs further.
\newline

The abundances in Table \ref{tbl1} cannot necessarily be directly compared to the solar system bodies. This is because the differential sinking times of Mn, Na, and Mg through the white dwarf's photosphere cause fractionation of the photospheric abundances away from that of the accreting material. For example, for Mn, Na, and Mg in the atmosphere of J0738+1835 the sinking times are 0.11, 0.17, and 0.18\,Myrs respectively \citep{Dufour2012}. The abundances tend from those of the accreted body (`build-up phase') to a steady state between accretion and diffusion on timescales of order the sinking timescale \citep{Koester09}. Once accretion has finished, abundances decrease in a `declining phase' \citep{Koester09}. In \cite{HarrisonThesis}, a Bayesian model is used to assess the most likely state of each body. The model finds that for GD362, J1535+1247, and WD\,0446-255 the accreting material is most likely in the build-up phase while J0738+1835 is likely to be in a steady state accretion phase.
\newline

\begin{table}[width=\linewidth,cols=4,pos=h]
\caption{The polluted white dwarf abundances (log number fraction)  adjusted from the values in Table \protect\ref{tbl1} in order to account for differential sinking. The abundances for GD362, J1535+1247, and WD\,0446-255 assume the polluting material is accreting in build-up phase (the abundances of the accreting material are identical to that of the atmosphere (Table \protect\ref{tbl1})) while the abundances for J0738+1835 assume that the polluting material is accreting in steady state (the abundances of the accreting material have been adjusted using the sinking timescales quoted in the text).}\label{tbl2}
\begin{tabular*}{\tblwidth}{@{} LLLLL@{} }
\toprule
System & log(Mn/Mg) & log(Mn/Na)\\
\midrule
GD362 & $-1.49 \pm 0.27$ & $+0.32 \pm 0.22$ \\
J0738+1835 & $-2.20 \pm 0.12$ & $-0.54 \pm 0.19$ \\
WD\,0446-255 & $-2.50 \pm 0.14$ & $-1.20 \pm 0.14$ \\
J1535+1247 & $-2.44 \pm 0.22$ & $-1.08 \pm 0.21$ \\
\bottomrule
\end{tabular*}
\end{table}

Table \ref{tbl2} outlines the expected abundance ratios of the rocky material that pollutes the white dwarfs GD362, WD\,0446-255, J1535+1247, and J0738+1835. These abundances can now be directly compared to those of the solar system rocky bodies.
\newline

\begin{figure}
	\centering
		\includegraphics[width=\columnwidth]{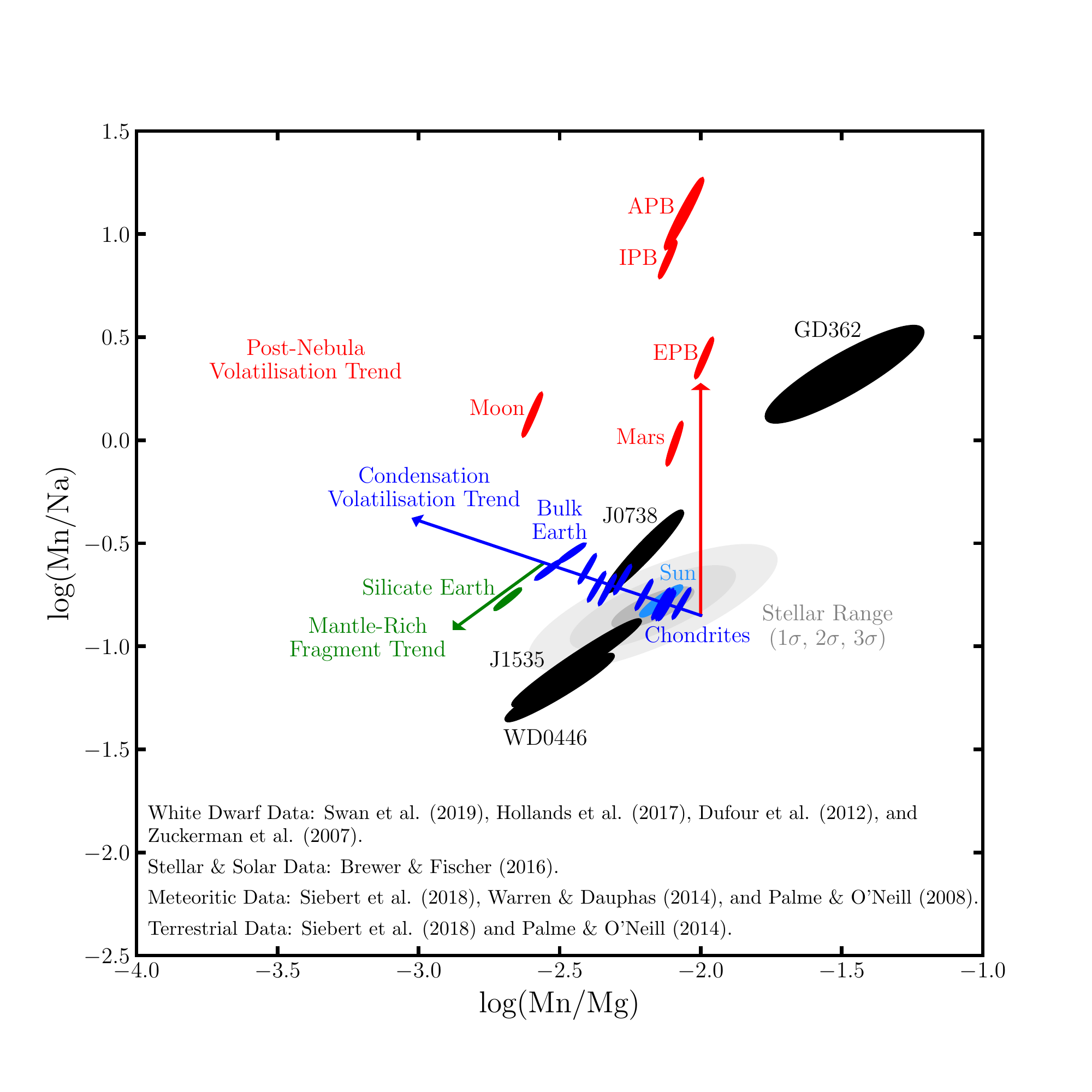}
	\caption{The Mn/Na and Mn/Mg ratios of the bodies which pollute white dwarfs (Table \protect\ref{tbl2}) are plotted onto Figure \protect\ref{FIG:1}. The white dwarf data (black) suggests that the three processes: condensation from the stellar nebula (blue), planetary differentiation (green), and post-nebula volatilization (red) may have occurred in exo-planetary systems, as the pollutant body ratios appear to have been moved along vectors away from the initial stellar abundance values (grey area). The errors are displayed as 1$\sigma$ error ellipses to capture the correlation between the axes.}
	\label{FIG:2}
\end{figure}

The abundance ratios from Table \ref{tbl2} of these four polluted white dwarfs are shown in Figure \ref{FIG:2}. An additional grey shaded region is added to account for the fact that in exo-planetary systems the initial composition of the stellar nebula may differ from that of the solar nebula. We model this potential variation using compositions estimated for nearby stars \citep{FischerBrewer2016}. If the stellar compositions found in \cite{FischerBrewer2016} represent the range of possible initial nebula abundances, then the possible nebula compositions are not likely to be sufficiently different from the solar system that the elemental condensation sequences derived for the solar system are invalidated. Crucially, there are no nearby stars with C/O ratios above the threshold at which carbide species preferentially condense out of the disc (C/O>0.75) and alter significantly the elemental condensation behaviour \citep{Moriarty2014}.
\newline

Figure \ref{FIG:2} highlights a diversity in the relative abundances of Mn, Mg, and Na in rocky exo-planetary bodies. The abundances of WD\,0446-255, J1535+1247, and J0738+1835 can be easily explained by condensation followed by differentiation and finally fragmentation of the body. This conclusion holds when all the derived abundances are analysed, not just Na, Mg, and Mn \citep{HarrisonThesis}. J1535+1247 is readily explained by the accretion of a primitive planetesimal with a composition identical to that of the stellar nebula from which it formed as it overlaps the grey shaded region in Figure \ref{FIG:2}, this is not only true for Mg, Mn, and Na but for all elements observed \citep{HarrisonThesis}. WD\,0446-255 has a composition which is consistent with the accreted material being a mantle rich fragment of a differentiated body which did not undergo substantial volatile loss during condensation. This is indicated by its position below grey-shaded region, readily accessible by core-mantle differentiation, as indicated by the green vector. These conclusions are reinforced in \cite{HarrisonThesis} where it is shown that the material has under abundances, relative to the stellar abundances, in all of the observed siderophiles (Fe, Ni, Mn etc.) and likely formed below 1000\,K. J0738+1835 is best explained by a scenario where the accreted material is a crust-stripped differentiated body which experienced limited volatile loss during condensation, as indicated by its high Mn/Na ratio, which lies above those predicted for initial nebula conditions (grey shaded region). This hypothesis holds when all observed elements are examined as there is a consistent depletion of the lithophiles (Ca, Mg, Si, Na etc.) in comparison to the stellar abundances \citep{HarrisonThesis}. A full discussion about polluted white dwarfs as evidence for exo-planetary differentiation and collisional processing is given in \cite{Harrison2018} and \cite{HarrisonThesis} and no further discussion will take place here.
\newline

Unlike the previously discussed objects, the Mn/Na abundance of GD362 is inconsistent with condensation volatilisation (>3$\sigma$). The abundances of the detected siderophile elements in GD362 (Fe, Ni, Cr etc.) are also inconsistent with the material being a fragment of a larger body which differentiated and was subsequently collisionally processed \citep{HarrisonThesis}. The stellar properties of GD 362 (shown in Table \ref{tbl3}) do not indicate any obvious differences from the other white dwarf stars considered, except in the unusually high level of trace hydrogen in its helium-rich atmosphere, a fact that initially made GD362 difficult to characterise \citep{Kawka2005}. However, this should not affect the abundance determinations used here as the models used in \cite{Xu2013} take this into consideration. The elemental abundances seen in GD362 are, therefore, difficult to explain without invoking post-nebula volatilisation to raise the Mn/Na ratio.
\newline

\begin{table}[width=\linewidth,cols=5,pos=h]
\caption{The stellar data for the four polluted white dwarfs analysed in this work. None of the major characteristics of GD362 are particularly unique in the sample analysed. The data is taken from \protect\cite{Zuckerman07,Dufour2012,Hollands2017,Swann2019}.}\label{tbl3}
\begin{tabular*}{\tblwidth}{@{} LLLLL@{} }
\toprule
System & Type & WD Mass & WD $\textrm{T}_{\textrm{eff}}$ & WD $\tau_{\textrm{Mg}}$\\
\midrule
GD362 & He & 0.72\,$M_{\odot}$ & 10,500\,K & 0.22\,Myrs \\
J0738+1835 & He & 0.84\,$M_{\odot}$ & 14,000\,K & 0.18\,Myrs \\
WD\,0446-255 & He & 0.58\,$M_{\odot}$ & 10,120\,K & 2.61\,Myrs \\
J1535+1247 & He & 0.60\,$M_{\odot}$ & 5,773\,K & 3.09\,Myrs \\
\bottomrule
\end{tabular*}
\end{table}

As the material that pollutes GD362 survived the giant branch evolution of the star, it is possible that devolatilisation could have occurred during the post-main sequence, when the luminosity of the star increases by many orders of magnitude, rather than shortly after or during formation. In order to investigate this possibility in Section 3 we model the compositional changes expected to occur to an asteroid in the GD362 system during the post-main sequence evolution of the host star.
\newline

\section{Modelling volatile loss during the post-main sequence}

When the progenitor of a white dwarf (a star with initial mass under 8-11 solar masses \citep{Siess2007}) evolves off the main sequence, it will go through phases where its luminosity exceeds 10,000 times the luminosity of the Sun and its radius exceeds a few astronomical units. During this post-main sequence evolution the equilibrium temperatures of the bodies orbiting the star can increase dramatically. Therefore, rocky bodies could potentially experience significant heating and volatile loss during this interval. 
\newline

\subsection{The conditions required to remove Na preferentially to Mn}

Our aim is to determine whether post-main sequence heating during the host star's giant branch evolution could heat an asteroid sufficiently to vaporise Na, whilst not vaporising Mn (or the whole asteroid for that matter), and produce the observed Mn/Na ratio in the pollutant body of GD362. First, we must estimate the sublimation temperatures of the Na and Mn species expected to be present in extra solar asteroids.
\newline

\begin{table}[width=\linewidth,cols=4,pos=h]
\caption{The possible solid, liquid, and gaseous species that could form when running the HSC Chemistry v. 8.0 equilibrium chemistry program to determine the behaviour of Na and Mn when heated. }\label{tbl4}
\begin{tabular*}{\tblwidth}{@{} LLLLL@{} }
\toprule
Gaseous Species & Liquid Species & Solid Species\\
\midrule
Na & Na & Na \\
\ce{Na2O} & \ce{Na2O} & \ce{Na2O} \\
\ce{Na2} & & \ce{NaO3} \\
NaO & & \ce{NaO2} \\
\ce{Na2O2} & & \ce{Na2O2} \\
O & \\
\ce{O2} & \\
\bottomrule
\toprule
Gaseous Species & Liquid Species & Solid Species\\
\midrule
Mn & Mn & Mn \\
MnO & MnO & MnO \\
\ce{MnO2} & \ce{Mn2O7} & \ce{MnO2} \\
O & & \ce{Mn2O3}\\
\ce{O2} & & \ce{Mn3O4} \\
\bottomrule
\end{tabular*}
\end{table}

In this work we used the software package HSC chemistry version 8 to produce vaporisation curves for Na and Mn. For Na we inputted 100\,kmol of solid \ce{Na2O} and 1000\,kmol of gaseous O into HSC chemistry and allowed it to equlibriate assuming the Na and O could only be in the form of the species listed in Table \ref{tbl4}. We tracked the percentage of Na in gaseous form as a function of temperature and pressure and the vaporisation curve was defined as the line in pressure temperature space at which over 10 percent of the Na was in the gas. For Mn we performed an analogous procedure, however in this case we inputted 100\,kmol of solid MnO and 1000\,kmol of gaseous oxygen into HSC chemistry and allowed it to equilibriate assuming the Mn and O could only be in the form of the species in Table \ref{tbl4}. We varied the abundances of the excess gaseous O in both cases to as low as 100\,kmol and found that this only caused the sublimation temperatures to vary by $\sim$50\,K and in both cases this variation caused the sublimation temperatures to increase. Therefore, variability in the abundance of available O will not dramatically effect our conclusions.
\newline

The vaporisation curves found are strong functions of pressure, therefore, in order to calculate the temperature at which Na starts to vaporise from an externally heated asteroid, one must know the pressure at which the potential vaporisation is occurring. Assuming that the atmosphere of the heated body is solely composed of the Na that is vaporised from its surface and the major factor contributing to atmospheric loss is Jeans loss, we can find the steady state mass of the atmosphere, and hence the surface pressure.

The steady state mass of the atmosphere $M_{\textrm{atmo}}$ is defined as
\begin{equation}
M_{\textrm{atmo}} = \Phi \tau_{\textrm{escape}}.
\end{equation}
The mass of Na sublimated per second, $\Phi$, is: 
\begin{equation}
\Phi = \frac{4 \pi R^{2}_{\textrm{ast}} \sigma \epsilon T^{4}}{C_{\textrm{Na}}},
\end{equation}
where $R_{\textrm{ast}}$ is the radius of the asteroid, $\sigma$ is the stefan-boltzmann constant, $\epsilon$ is the emissivity of the asteroid, $T$ is the surface temperature, and $C_{\textrm{Na}}$ is the latent heat of vaporisation of Na.

The Jeans escape timescale, $\tau_{\textrm{escape}}$ is given by
\begin{equation}
\tau_{\textrm{escape}} = \sqrt{\frac{2 \pi k R^{4}_{\textrm{ast}} T}{G^{2} M^{2}_{\textrm{ast}} \mu}} \frac{e^{\lambda}}{(1+\lambda)},
\end{equation}
where
\begin{equation}
\lambda = \frac{G M_{\textrm{ast}} \mu}{k R_{\textrm{ast}} T},
\end{equation}
$k$ is the boltzmann constant, $M_{\textrm{ast}}$ is the mass of the asteroid, $G$ is the gravitational constant, and $\mu$ is the mean molecular mass of the atmosphere.

As the gravitational surface pressure is defined as
\begin{equation}
P = \frac{G M_{\textrm{ast}} M_{\textrm{atmo}}}{4 \pi R^{4}_{\textrm{ast}}},
\end{equation}

\noindent surface pressure can be written as a function of asteroid surface temperature, asteroid radius and asteroid mass:
\begin{equation}
P = \sqrt{\frac{2 \pi k \sigma^{2} \epsilon^{2} T^{9} }{\mu C_{\textrm{Na}}^{2}}} \frac{e^{\frac{G M_{\textrm{ast}} \mu}{k R_{\textrm{ast}} T}}}{(1+\frac{G M_{\textrm{ast}} \mu}{k R_{\textrm{ast}} T})}.
\end{equation}

\begin{figure}
	\centering
		\includegraphics[width=\columnwidth]{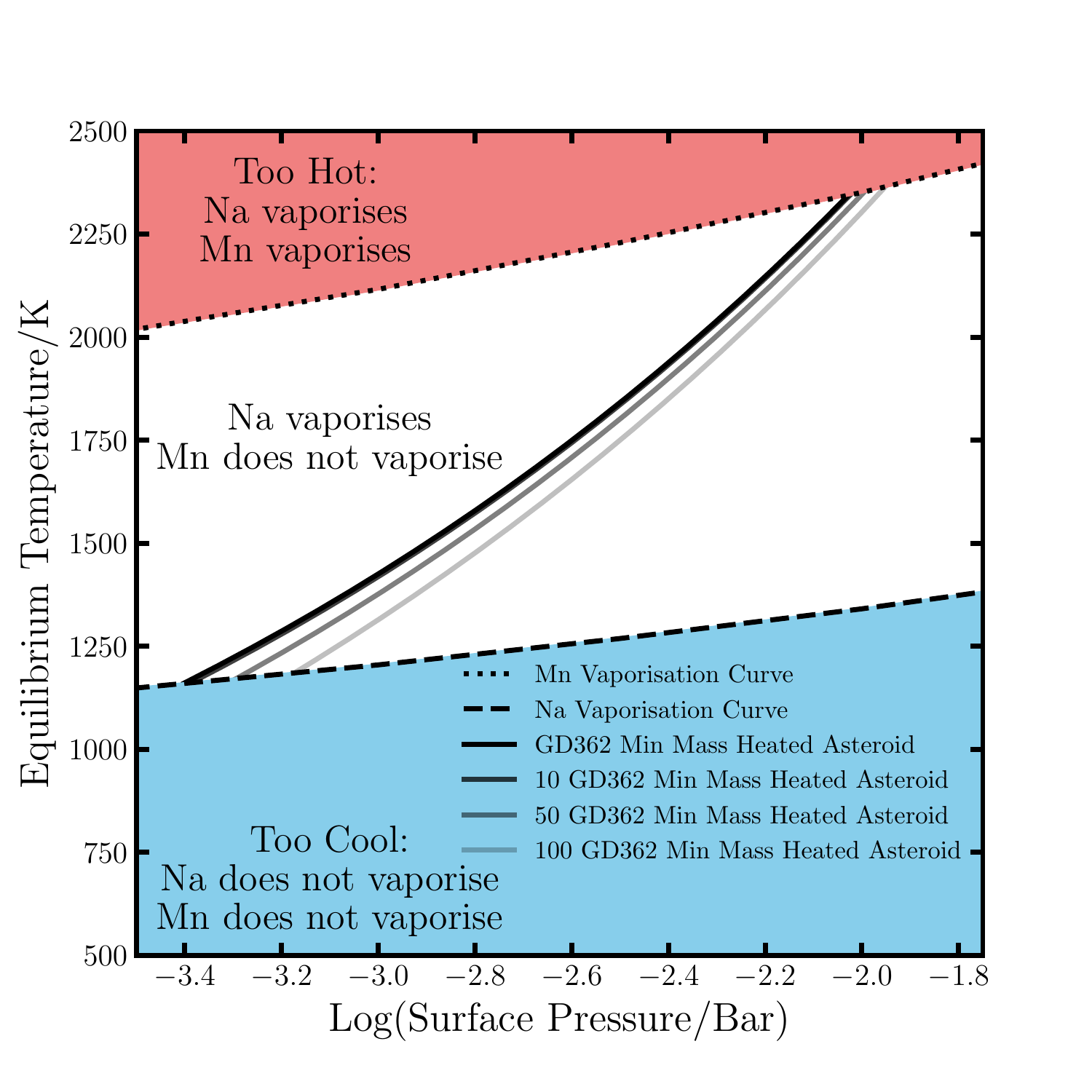}
	\caption{The solid lines show the equilibrium temperature of an externally heated asteroid as a function of surface pressure and asteroid mass. The white region indicates the region of interest where Na vaporises but Mn does not. The Mn and Na vaporisation curves were found using the software package HSC Chemistry version 8 (For further details see Section 3.1).}
	\label{FIG:3}
\end{figure}

Figure \ref{FIG:3} shows how for externally heated spherical black body asteroids, of density 3000\,$\textrm{kg}\,\textrm{m}^{-3}$ and Bond albedo 0.035, the surface pressure varies with surface temperature and asteroid mass. The body is assumed to have no initial atmosphere and the steady state atmosphere produced is only composed of vapourised Na. The four solid lines shown correspond to four different asteroid masses and highlight how larger asteroids retain larger atmospheres and, therefore, require larger temperatures to cause vaporisation. We find that in the minimum mass scenario (where we assume all of the asteroid is currently in the white dwarf's atmosphere and none is left in a disc around the star, the value used is $6.31 \times 10^{19}$\,kg \citep{Xu2013}) for temperatures above 1186\,K, and below 2344\,K, Na will vaporise from the surface while Mn will not. We also find that even if the mass of the asteroid polluting GD362 is 100 times larger than the minimum mass scenario, the above quoted temperatures vary by less than 40\,K.
\newline

\subsection{Can post-main sequence heating remove Na preferentially to Mn?}

The evolution of the luminosity of the progenitor to GD362 is calculated using the single star evolution (SSE) code \citep{SSE2013}. The code was run assuming an initial stellar mass of 3.2 solar masses. The mass of the white dwarf GD362 is 0.72 solar masses \citep{Xu2013}. The value of 3.2 solar masses was chosen because it is the initial stellar mass which, given the star is solar metallicity, results in the formation of a white dwarf of 0.72 solar masses \citep{WDMass}. The stellar luminosity calculated can then be converted into an equilibrium temperature which is a function of radial distance from the star and can then be compared to the required vaporisation conditions. However, not all bodies in the planetary system will survive until the white dwarf phase. Bodies with close-in orbits (small radial distances from the star) can be either engulfed by the star as it expands or be spun to break up as the star's luminosity increases.
\newline

The Yarkovsky–O'Keefe–Radzievskii–Paddack (YORP) effect causes asteroids to become spun up by stellar radiation \citep{YORP}. On the giant branches, when the star's luminosity greatly increases, asteroids may be spun up to the point of break up \citep{VerasYORP}. Smaller bodies which are closer to the star are easier to destroy due to the YORP effect. Therefore, for GD362, the YORP effect would be maximised for the case where the polluting asteroid only has mass equal to that of the material in the atmosphere (minimum mass assumption). The code presented in \cite{VerasYORP} calculates that if the body was interior to 0.4\,au it would be spun to break up during the giant branch, assuming the body was the minimum possible mass.
\newline

During post-main sequence evolution, the radii of a star can increase by orders of magnitude, potentially causing bodies which orbit too close to the star to become engulfed and destroyed. Whilst strong stellar winds during this phase of radial expansion cause the orbits of bodies around the star to migrate outwards as the star loses mass, this is often not enough to stop engulfment \citep{Mustill2012, Adams2013}. Using the analytical expression given in \cite{Adams2013}, for the minimum mass case for GD362, we find that the body would be engulfed inside of 0.46\,au. The parameter values inputted into the analytical expression were those given by the SSE code for a 3.2 solar mass star: An initial AGB stellar mass of 3.2 solar masses, an initial AGB stellar radius of 1.22\,au, and an AGB duration time of 0.96\,Myrs. The gamma parameter was conservatively set to a value of 1, in order to minimise the engulfment radius. Larger bodies are engulfed more readily, and therefore, need to orbit further from the star in order to avoid destruction \citep{Mustill2012, Adams2013}. Thus, for all possible  masses of the pollutant of GD362, engulfment will be the dominant factor in determining the closest orbit the body could have been on and survived to the white dwarf phase. Therefore, in order to consider the case of maximum heating, and therefore maximum Na loss, we assume the minimum possible mass for the progenitor to the GD362 pollutants, as it allows the body to orbit closer to the star without being destroyed. We note here that this assumption is somewhat contradictory to the assumption that the white dwarf is currently accreting material (i.e., that the mass of the object is large enough to have supplied all material seen in the white dwarf's atmosphere and still provide a larger reservoir of as yet unaccreted material.). However, it is consistent with our attempt to model the maximum possible fractionation of Mn from Na that could have occurred during the post-main sequence life of the star, to see whether earlier post-nebula volatilisation can be ruled out.
\newline

\begin{figure}
	\centering
		\includegraphics[width=\columnwidth]{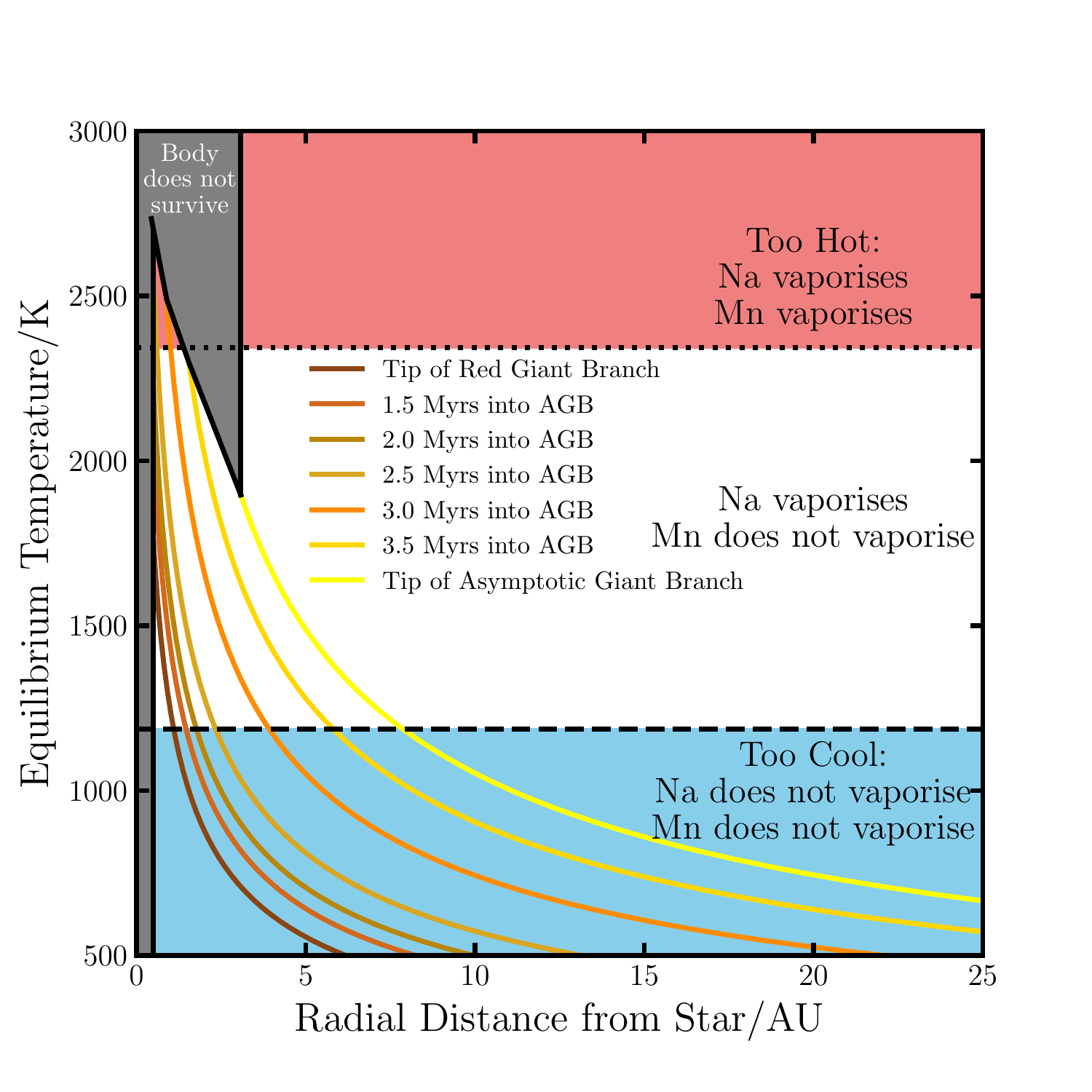}
	\caption{The solid lines show the equilibrium temperature of a spherical black body asteroid as a function of radial distance from an initially 3.2 solar mass star at various epochs which are 0.5\,Myrs apart. If the equilibrium temperatures enter the white region Na can be vaporised from the surface while Mn is retained. The grey area is the region for which an asteroid of density 3000\,$\textrm{kg}\,\textrm{m}^{-3}$ and radius 170\,km would be destroyed by stellar engulfment which in this scenario dominates over the YORP effect \protect\citep{Adams2013, VerasYORP}. For further details see Section 3.2.}
	\label{FIG:4}
\end{figure}

Figure \ref{FIG:4} displays the equilibrium temperature of a spherical black body asteroid of Bond albedo 0.035 orbiting a 3.2 solar mass star as a function of radial distance from the star and time. The luminosity as a function of time was calculated using the SSE code \citep{SSE2013}. The grey area is the region for which an asteroid of density 3000\,$\textrm{kg}\,\textrm{m}^{-3}$ and radius 170\,km (the minimum mass assumption) would be destroyed by stellar engulfment \citep{Adams2013}. The regions where Mn and Na vaporise were taken from Figure \ref{FIG:3} assuming the minimum mass scenario. By assuming the minimum mass scenario we minimise the size of the grey area, thus, maximising the size of the region where Na can be vaporised from the surface while Mn is retained.
\newline

Figure \ref{FIG:4} highlights how the surface of an asteroid of radius 170\,km orbiting the progenitor of the star GD362 would be at temperatures such that Na vaporises while Mn would not for potentially up to 4 million years. Figure \ref{FIG:4} additionally displays that the vaporisation of Na could occur on all bodies interior to approximately 8\,au.
\newline

\subsection{Can post-main sequence heating produce a significant change to a body's Mn/Na ratio?}

In order to calculate whether a sufficient fraction of the Na from the body can be vaporised and lost, thereby, altering the bulk composition of the asteroid, we must calculate how the temperature of the interior of the asteroid evolves during the post-main sequence.
\newline

Thoroughly investigating this would require a complete asteroid interior model. However, due to the uncertainty on the mass of the asteroid and the abundances in this work, we adopt a simple heat diffusion model in order to calculate a maximum possible volume heated and, therefore, the maximum fraction of Na lost.
\newline

In order to calculate the temperature of the interior at a given depth and at a given time we assume the asteroid is a sphere. Therefore, the relevant heat diffusion equation is:

\begin{equation}
\frac{\partial T}{\partial t} = \frac{\kappa}{\rho C_{\textrm{b}}} \frac{1}{r^{2}} \frac{\partial}{\partial r}\left( r^{2} \frac{\partial T}{\partial r} \right),
\end{equation}

\noindent where $T$ is the temperature of the asteroid at a given distance from the centre $r$ and a given time $t$, $\kappa$ is the thermal conductivity of the asteroid, $\rho$ is the density of the asteroid, and $C_{b}$ is the heat capacity of the asteroid.

The boundary conditions for an isothermal sphere of radius a are:

\begin{equation}
T(a,t) = T_{1} \, \, \, \, \, \, \, \, \, \, T(r,0) = T_{0}.
\end{equation}

Assuming a steady state is reached the solution must have the form

\begin{equation}
T = A + \frac{B}{r},
\end{equation}

\noindent therefore, the following substitution can be used

\begin{equation}
B(r,t) = r(T(r,t)-T_{1}).
\end{equation}

\noindent Equation 7 then becomes

\begin{equation}
\frac{\partial B}{\partial t} = \frac{\kappa}{\rho C_{\textrm{b}}} \frac{\partial^{2} B}{\partial r^{2}},
\end{equation}

\noindent and the boundary conditions become

\begin{equation}
B(a,t) = 0 \, \, \, \, \, B(r,0) = r(T_{0}-T_{1}) \, \, \, \, \, B(0,t) = 0.
\end{equation}

\noindent The solution to Equation 11 is

\begin{equation}
B(r,t) = \frac{2a}{\pi} (T_{1}-T_{0}) \sum_{n=1}^{\infty} \frac{(-1)^{n}}{n} sin \left(\frac{n \pi r}{a} \right) e^{-\frac{\kappa n^{2} \pi^{2} t}{\rho C_{\textrm{b}} a^{2}}},
\end{equation}

\noindent therefore the solution to Equation 7 is

\begin{equation}
T(r,t) = T_{1} + \frac{2a}{\pi r} (T_{1}-T_{0}) \sum_{n=1}^{\infty} \frac{(-1)^{n}}{n} sin \left(\frac{n \pi r}{a} \right) e^{-\frac{\kappa n^{2} \pi^{2} t}{\rho C_{\textrm{b}} a^{2}}}.
\end{equation}

This solution can then be used to calculate the temperature at a given depth inside the asteroid during the post-main sequence evolution of the star. In this work, we make the simplistic assumption that if part of the asteroid gets to the temperature required to vaporise Na on the surface, then it is possible for the Na to be lost to this depth. This will certainly be an overestimate as, firstly, the temperature required to vaporise Na will actually be higher the deeper in the asteroid due to the increased pressure and, secondly, being at depth inside the asteroid will make it more difficult for the Na to out-gas and leave the asteroid once it does vaporise.
\newline

The distribution of Na in the asteroid may not be uniform. Hence, if we wish to find the fraction of Na it is possible to heat, we need to estimate the fraction of Na at a given depth. In this work we calculate two end member assumptions. Firstly, a case where Na is distributed homogeneously throughout the whole body. This is analogous to the asteroid being primitive and undifferentiated. Secondly, a case where Na is mainly sequestered in the upper layers of the asteroid. This is analogous to the asteroid being differentiated into a core, a mantle, and a crust. We assume the maximum size core a body of radius 170\,km could differentiate into is 85\,km and the maximum thickness of the crust is 6\,km. 6\,km was chosen, as if we fix the composition of this crust to be the same as the Earth's oceanic crust \citep{White2014}, at 6\,km thickness the interior would be depleted in an incompatible manner. Assuming no Na is sequestered into the core and the crust has the same composition as the Earth's oceanic crust then 17\% of the body's Na is in the first 6\,km and 83\% of the body's Na is in the next 79\,km.
\newline

\begin{figure}
	\centering
		\includegraphics[width=\columnwidth]{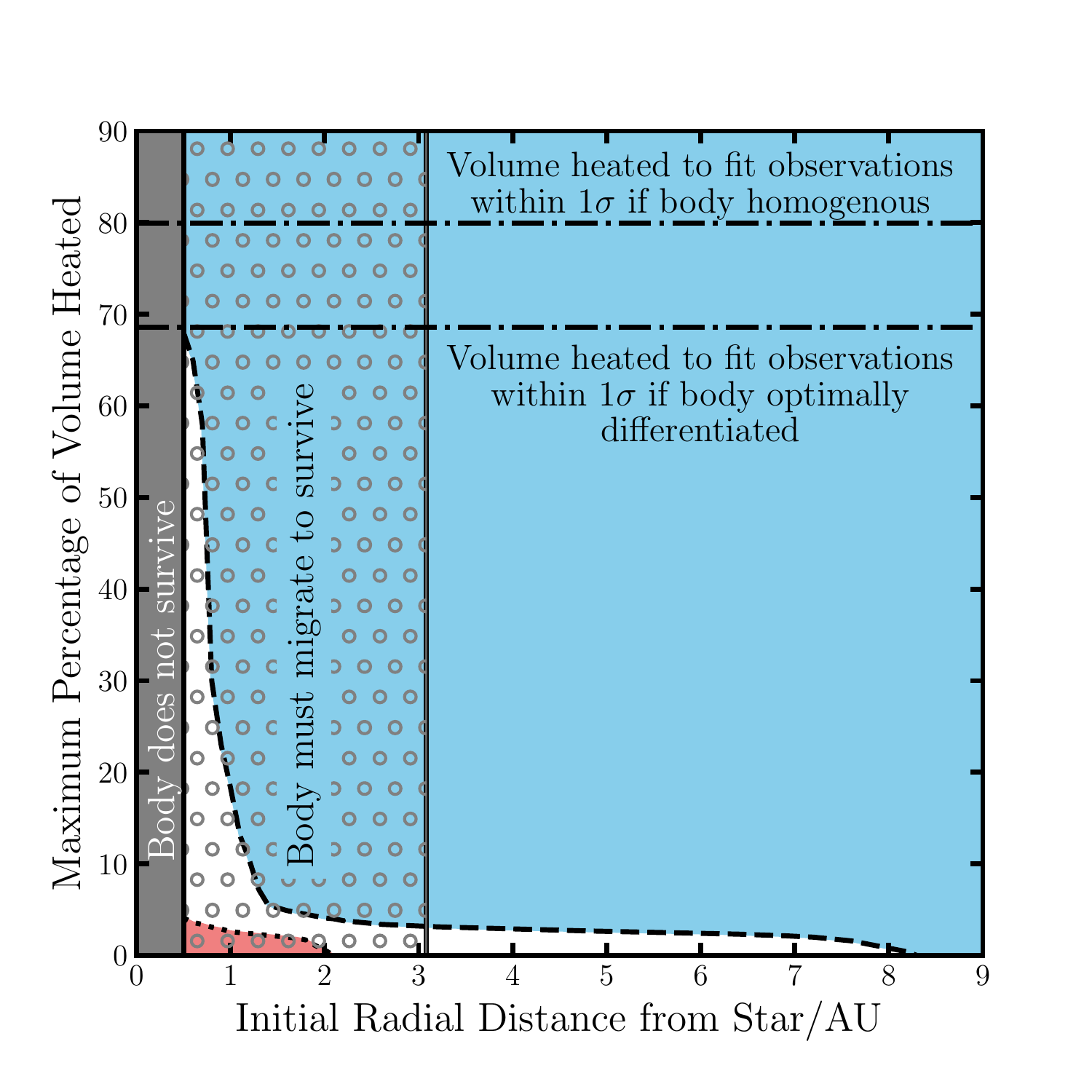}
	\caption{The maximum percentage of the volume of a spherical 170\,km radius 3000\,$\textrm{kg}\,\textrm{m}^{-3}$ density asteroid which is heated to a given temperature as a function of initial radial distance from the star. The white area is the region where the asteroid is heated to a temperature such that Na on the surface would vaporise while Mn on the surface would not. The reason this plot highlights the maximum possible percentage heated is because the calculated percentage of the volume heated assumes a fixed radial location whereas in reality due to stellar mass loss all bodies will migrate outward over time.}
	\label{FIG:5}
\end{figure}

Figure \ref{FIG:5} shows the maximum percentage of volume heated for the minimum mass assumption asteroid during the post-main sequence evolution of GD362, as a function of initial radial distance from the star. We assume a heat capacity of 840\,$\textrm{J}\,\textrm{kg}^{-1}\,\textrm{K}^{-1}$ and a thermal conductivity of 2\,$\textrm{J}\,\textrm{s}^{-1}\,\textrm{m}^{-1}\,\textrm{K}^{-1}$. As outlined earlier, as orbital migration occurs during the evolution of the star the radial locations in Figure \ref{FIG:5} are the initial orbital distances and, therefore, the percentage of volume heated is an upper bound. The volume required to be heated to fit the abundance observed in GD362 to within the 1$\sigma$ uncertainties depends on the distribution of Na in the body. Figure \ref{FIG:5} shows the volumes required for a homogeneous body and a maximally differentiated one: it is very difficult to heat a large fraction of the body on the post-main sequence. Thus, it is not possible to produce the observed abundances to within their 1$\sigma$ uncertainties. This is mainly due to the timescales of heat diffusion being longer than the time it takes to evolve through the post-main sequence evolution. In fact, Figure \ref{FIG:5} shows that any body orbiting outside of the first astronomical unit will only experience heating to a very small fraction of the body. Our result is consistent with previous work that has noted the difficulty of post-main sequence heating to drastically change the composition of asteroidal bodies, for example \cite{Malamud2016} and \cite{Malamud2017a} showed that even interior ice species are expected to survive the giant branch. 
\newline

\begin{figure}
	\centering
		\includegraphics[width=\columnwidth]{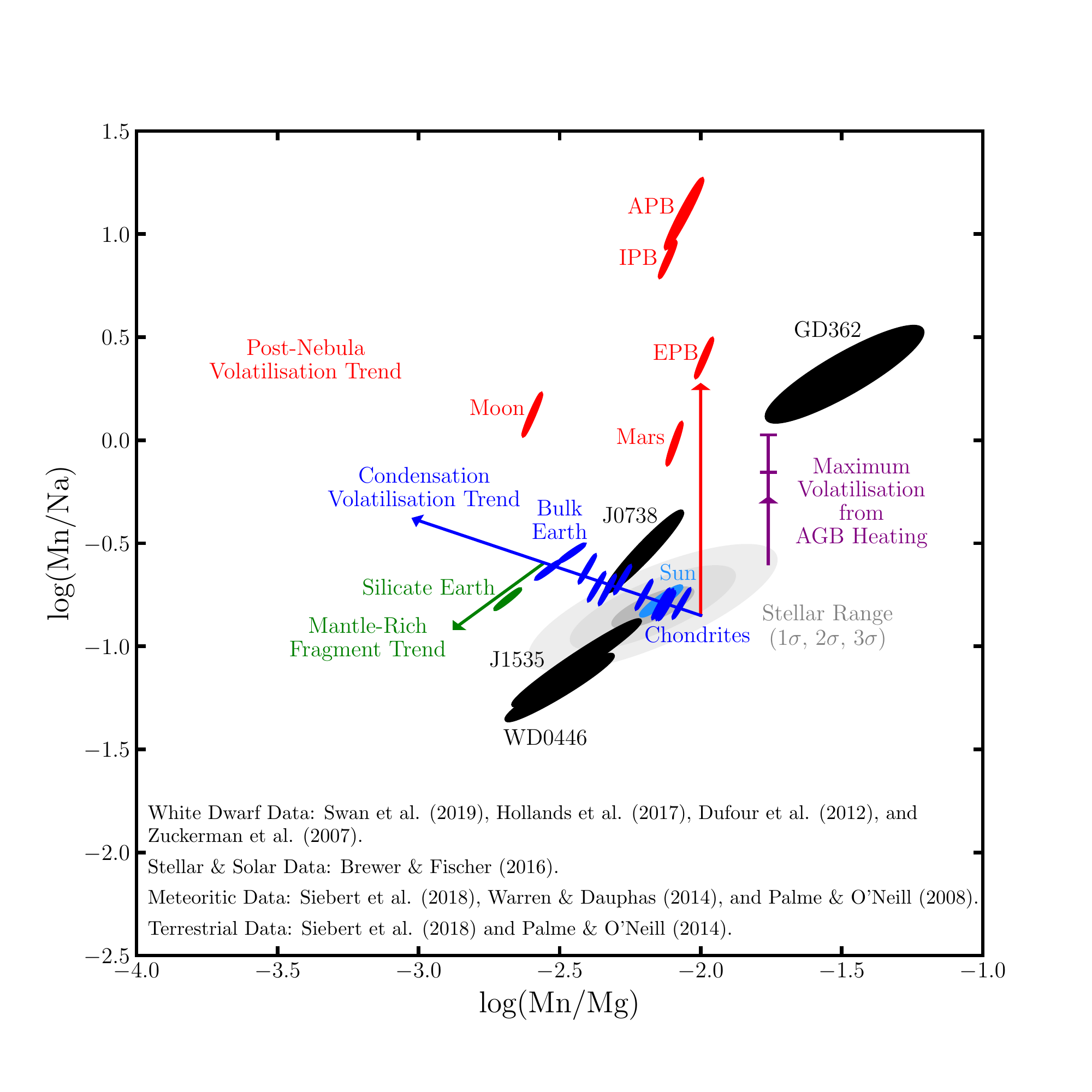}
	\caption{An updated version of Figure \protect\ref{FIG:2} which includes two additional vectors (purple) which indicate the maximum possible change in composition which could be attributed to heating during the post-main sequence evolution of the star. The shorter vector assumes the asteroid is homogeneous, whereas, the longer vector assumes a scenario where Na is easier to remove due to the fact it is mainly sequestered in the upper layers of the asteroid. For further details see Section 3.3.}
	\label{FIG:6}
\end{figure}

Figure \ref{FIG:6} is an updated version of Figure \ref{FIG:2} now including the maximum possible change in Mn/Na estimated from our work due to post-main sequence heating. Two additional purple vectors are plotted; the longer vector assumes the body is differentiated, whereas the shorter vector assumes the body is homogeneous. We emphasise that these vectors are upper limits for what post-main sequence heating could achieve in terms of Mn/Na fractionation, as we assume the mass of the pollutant is the minimum possible mass, therefore, maximising both its volume heated and the possible proximity to the star whilst avoiding engulfment.
\newline

The results of Figure \ref{FIG:6} highlight that it is not possible to produce the abundance pattern seen in the pollutant of GD362 to within the quoted 1$\sigma$ error bars with post-main sequence stellar heating.

\section{Discussion}

The main aim of this work was to investigate how volatiles are lost in exo-planetary bodies. In the solar system, Mn and Na abundances suggest that there are three key processes: nebula condensation, differentiation and fragmentation, and post-nebula volatilisation. In order to probe these effects in exo-planetary systems we utilise the Mn, Mg, and Na abundances of the planetary material that has accreted onto white dwarfs. Application of this method to exo-planetary systems does come with additional assumptions and complications, most notably an extra potential phase of volatile loss when the star expanded on the giant branch before becoming a white dwarf. Evidence from the four white dwarfs investigated suggests that only condensation, differentiation and post-nebula volatilisation are required, and our simple model rules out heating on the giant branches as a likely explanation for the observed Mn/Na abundance of GD362. We now discuss the validity of our results and justify the assumptions made in our simple post-main sequence heating model.
\newline

\subsection{Model assumptions}

In order to discuss the validity of our results we must first address the validity of white dwarfs as a `laboratory' to study the composition of exo-planetary material. Currently, white dwarfs offer a unique insight into the bulk composition of exo-planetary bodies, which can not be offered by the study of exo-planet masses and radii or atmospheric compositions. As discussed in Section \ref{PWDD} the prevailing explanation for the presence of metals in cool white dwarf atmospheres is the accretion of exo-planetary material. In this work we assume that the metals originate from one body and that they can be related to the elemental abundances of the exo-planetary body accreted via consideration of the relevant sinking timescales. It is possible that the white dwarfs are polluted by multiple bodies. However, even if this was the case the metal abundances will be dominated by the largest body, meaning the assumption of a single accreting body may still effectively hold in this case. Even if multiple similar mass bodies did pollute GD362, at least one of the bodies would be required to have an enhanced Mn/Na ratio and such a scenario would make the observed abundance an underestimate of the true Mn/Na fractionation that body experienced. It is also possible that GD362 is not accreting material in the build-up phase and the abundances in the atmosphere are related to the abundances of the accreting material in a manner dissimilar to those calculated in this work. However, as Mn sinks faster than Na, if GD362 was in fact in a different phase of accretion the true Mn/Na ratio of the original body would in fact be even higher. Therefore, we do not expect our assumptions regarding the link between polluted white dwarf atmospheres and exo-planetary compositions to affect our results.
\newline

Only four white dwarf systems have measurements of the elements required to probe the nature of planetary volatile loss (Mg, Mn, and Na). The lack of observations are due to the fact that only sufficiently heavily polluted white dwarfs that lie in a narrow temperature range produce strong enough Mn and Na absorption features for both elements to be simultaneously detected. Therefore, in order to robustly conclude that only condensation, differentiation and post-nebula volatilisation are required to explain the abundances in exo-planetary material, additional observations of heavily polluted white dwarf systems within the relevant temperature range will be required. 
\newline

The major caveats which affect the validity of our results involve the simple model we established in order to rule out post-main sequence heating as an alternative explanation for the enhanced Mn/Na ratio of GD362. The major caveats of the model can be separated into two categories. Caveats which have been designed specifically so that they maximise the post-main sequence heating effect, and caveats which do not, and therefore, could potentially affect the conclusions of this work. The following assumptions are ones which could potentially alter the conclusions of this work:
\begin{itemize}
\setlength\itemsep{1em}

\item \emph{Modelling the post main-sequence}: The duration of the giant branches, the maximum luminosity on the giant branches, and the closest possible orbit for which a body can survive to the white dwarf phase are all functions of the initial main sequence mass of the star. A longer duration for the AGB phase would lead to increased heating and could potentially explain the observed signature. For GD362 to spend significantly longer on the giant branch than we have modelled, given its observed mass of $0.72\pm0.02$ solar masses \citep{Xu2013}, GD362 would need an initial metallicity higher than 3 times solar or lower than one third solar, in either of these cases, GD362 would have an initial mass which was considerably less than than 3.2 solar masses and therefore spend longer on the giant branch \citep{WDMass}. However, the majority of nearby stars do not have metallicities which are this extreme, therefore, it is unlikely that the metallicity of GD362 could be such that our conclusions would be affected \citep{FischerBrewer2016}. Additionally, if the model used to calculate the progenitor mass was inaccurate this could also cause the AGB lifetime to extend. However, while there is uncertainty on the progenitor masses predicted by each model, the masses predicted by various different methods \citep{ElBadry2018, Cummings2018} are consistent with the mass used in this work.

\item \emph{Modelling Na loss}: In this work we assumed chemical equilibrium would be reached when calculating the sublimation temperature of Na and Mn and that sublimation would occur in an environment with plentiful oxygen. It is not obvious whether these assumptions will hold and indeed the volatilisation of Na and Mn as a function of prevailing conditions is complex and the basis of ongoing work \citep{Sossi2018}. Additionally, the volatilisation will occur to the substances when they are in a complex multi-component silicate melt, where the activity of the Mn and Na components will play a vital role in the volatilisation process. However, regardless of the exact temperature required to vaporise Na, unless it becomes substantially easier to vaporise Na, it will remain difficult to heat a large enough volume of the body to create the observed elevated Mn/Na ratio. In any case, the activities of Na and Mn in silicate liquids reported in \cite{Sossi2018} suggest that our simple model will in fact overestimate the ease at which Na is lost rather than underestimate it.

\item \emph{Modelling silicate vapour escape}: In order to calculate the surface pressure of the body it is assumed that the vaporised Na escapes the planetary body via Jeans escape. In reality many escape mechanisms may be at work, for example the hydrodynamic and sputtering escape mechanisms. If these escape mechanisms are important, their efficiency will cause a reduction in surface pressure, which will decrease the vaporisation temperature of Na and will potentially allow it to be more readily lost. However, this decrease is not expected to be drastic enough to change our conclusions as Na will still be difficult to vaporise from deep inside the body's interior.
\end{itemize}

The following assumptions have been designed such that they maximise the potential for the abundances in GD362's atmosphere to be explained by post-main sequence heating:

\begin{itemize}
\setlength\itemsep{1em}
\item The mass of the polluting body is equal to the total mass of the metals in the atmosphere of GD362. Less massive bodies can survive closer in orbits and can have more of their total volume heated in a given time, therefore, minimising the mass of a body maximises the potential heating it can experience.
\item The pollutant body has an atmosphere solely composed of material which sublimates from its surface. This lowers the total surface pressure and therefore the sublimation temperatures.
\item The chosen model parameters for the asteroid survival models presented in \cite{Adams2013} and \cite{VerasYORP} have been set such that the destruction distances for planetesimals during the evolution of the star are minimised.
\item The pollutant body can survive on a stellar-surface-skimming orbit. 
\item Any Na which reaches the sublimation criteria can escape from the polluting body.
\item The pollutant can be a differentiated body with a thick crust and a core that has a radius of half that of the body, allowing the majority of the Na to be sequestered in the upper layers of the pollutant.
\end{itemize}

\subsection{Processes during planet formation as traced  by polluted white dwarfs}

This work has shown that the Mg, Na, and Mn abundances of three of the analysed polluted white dwarf systems can be well explained by condensation and differentiation processes. This provides evidence that the main processes which determined the bulk composition of the rocky worlds in the solar system have determined the bulk composition of the rocky worlds in exo-planetary systems. This reinforces previous findings which have suggested that the geological processes which occur in white dwarf planetary systems do not appear to be dissimilar from the geological processes which occur in the solar system \citep{JuraYoung2014, Harrison2018, Doyle2019, Doyle2020}.
\newline

The most significant result of this work is that one polluted white dwarf system, GD362, requires post-nebula volatilisation. GD362 is a historically significant system as it was the first polluted white dwarf to have the abundances of the metals in its atmosphere measured in detail \cite{Zuckerman07}. Additionally, GD362 has abundance measurements of 16 different metal elements, the most of any single system to date \citep{Zuckerman07,Xu2013}. The differences between the depletion of volatiles due to incomplete condensation from the nebula, and the depletion of volatiles due to post-nebula heating are well understood \citep{ONeillPalme2008,Visscher2013,Siebert2018}. The enhanced Mn/Na ratio of GD362 is a signature of post-nebula volatilisation and, crucially, we find that the required volatilisation cannot be easily produced during the post-main sequence evolution of the star. This implies that the planetesimal which pollutes GD362 underwent a period of heating such that it developed a global magma ocean once the stellar nebula had dissipated.
\newline

The formation of global magma oceans is predicted to have occurred on many of the solar system's minor bodies. The heat required to form global magma oceans is expected to be generated by a combination of impact heating from planetary collisions and short-lived radionuclides. This heating must have been able to melt the deep interior of the planetesimal in order to remove the required fraction of Na. The feasibility of this mechanism is not calculated in this paper, however, we note that the position of GD362 on Figure \ref{FIG:6} is not dissimilar to that of Vesta (the Eucrite parent body, EPB). Therefore, it seems likely that the same process predicted to have removed volatiles from Vesta could devolatilise the pollutant of GD362. Additionally, we note that the availability of short-lived radionuclides in exo-planetary systems is not well constrained and whether or not short-lived radionuclides pollute exo-planetary systems remains a subject of debate due to their stochastic production process \citep[e.g.][]{Lichtenberg2016,Young2016}. It is also possible that the heating process could have been powered by collisions alone, either way, additional white dwarf observations capable of identifying post-nebula volatilisation could potentially provide useful insights and constraints into this area of research.
\newline

The Mn/Mg ratio of GD362 is potentially difficult to explain. As highlighted by Figure \ref{FIG:2}, it is simply possible that the progenitor star had an unusually low Mg abundance. However, analysis of all the elements present in the atmosphere given in \cite{HarrisonThesis} found that the estimated Mg value for the pollutant of GD362 is potentially too low, and that re-observation would likely yield a higher abundance measurement. Additional observations of polluted white dwarf systems could yield more systems with enhanced Mn/Na ratios and add further weight to the conclusions presented here, and therefore, would be a worthwhile project. Additionally, further modelling to investigate whether post-main sequence heating can contribute to smaller changes in the Mn/Na ratio of planetary bodies would be of great interest, especially once more white dwarf systems with Mn and Na abundances are discovered.
\newline

\section{Conclusions}

Volatile loss is a key process in rocky planetary bodies. Mn and Na trace the loss of volatiles and, crucially, can distinguish between volatile loss occurring under two physical-chemical regimes. First, the incomplete condensation of the nebula gas, early in a system's evolution. Second, the loss of volatiles late in a system's evolution, after the nebula gas has dissipated.
\newline

The Mn to Na ratio and Mn to Mg ratio observed in the material accreted by polluted white dwarfs can be used to provide evidence for condensation processes and post-nebula volatilisation occurring in exo-planetary systems. In this study we found that the abundances present in the material polluting J0738+1835, J1535+1247, and WD\,0446-255 are consistent with a scenario in which the material condensed out of a protoplanetary disc, before undergoing differentiation and collisional processing, and then finally accreting onto the white dwarf. The Mn/Na ratio of the material polluting the star GD362 cannot be explained by condensation volatilisation processes (>3$\sigma$). We hypothesise that the enhanced Mn/Na ratio is a signature of post-nebula volatilisation. We show that any alterations to the composition of the material orbiting GD362, that could develop due to heating during the giant branch evolution of the star, are not significant enough to increase the Mn/Na ratio to match that observed in GD362. Even if the polluting body was suitably small, suitably differentiated, and orbited its host star on a surface grazing orbit we still cannot explain the abundances to within their 1$\sigma$ error bars. Therefore, we conclude it is most likely that the volatile loss that occurred on the pollutant of GD362 after the dissipation of the nebula gas was due to impact heating and/or short-lived radionuclides, which created a global magma ocean, allowing Na to be efficiently outgassed, similar to the process experienced by small rocky bodies of the solar system. Therefore, GD362 may provide evidence for the occurrence of global magma oceans and post-nebula volatilisation in exo-planetary systems.

\section{Acknowledgements}

We would like to thank Dimitri Veras for his helpful, useful, and insightful comments regarding post-main sequence stellar evolution. We would also like to thank Mark Hollands for providing additional data for the white dwarf system SDSSJ1535+1247. We also thank Mark Wyatt for his useful comments which improved the quality of the paper.  We are also grateful to the Science \& Technology Facilities Council, and the Royal Society - Dorothy Hodgkin Fellowship for funding the authors of this paper. We would also like to thank the anonymous reviewers whose comments helped improve this manuscript.

\bibliographystyle{cas-model2-names}

\bibliography{cas-refs}

\end{document}